\documentclass[article,reprint,amsmath,amssymb,superscriptaddress,longbibliography]{revtex4-1}%
\usepackage{graphicx}
\usepackage{dcolumn}
\usepackage{bm}

\usepackage[colorlinks,
            linkcolor=blue,
            anchorcolor=blue,
            citecolor=blue,
            ]{hyperref}
\usepackage{breakurl}
\usepackage{multirow}
\usepackage{enumitem}
\usepackage{tablefootnote}
\usepackage{threeparttable}
\usepackage{tabularx}
\usepackage{ulem}
\setcitestyle{super}

\usepackage{lineno}

\begin{document}

\title{Phase engineering of giant second harmonic generation in Bi$_2$O$_2$Se}
	

\author{Zhefeng Lou$^{1,2,3,\dag}$, Yingjie Zhao$^{4,\dag}$, Zhihao Gong$^{5,6,\dag}$, Ziye Zhu$^{4}$, Mengqi Wu$^{4}$, Tao Wang$^{1}$, Jialu Wang$^{7,1}$, Haoyu Qi$^{8,1}$, Huakun Zuo$^{9}$, Zhuokai Xu$^{1,2,3}$, Jichuang Shen$^{4}$, Zhiwei Wang$^{8,10,11}$, Lan Li$^{4}$, Shuigang Xu$^{1,3}$, Wei Kong$^{4}$, Wenbin Li$^{4,*}$, Xiaorui Zheng$^{4,*}$, Hua Wang$^{5,12,*}$ $\&$ Xiao Lin$^{1,3,*}$
~\\
	{\small\textit{$^1$Key Laboratory for Quantum Materials of Zhejiang Province, Department of Physics, School of Science and Research Center for Industries of the Future, Westlake University, Hangzhou 310030, P. R. China. 
	\\ 
    $^2$School of Physics, Zhejiang University, Hangzhou 310027, Zhejiang Province, China.
    \\
    $^3$Institute of Natural Sciences, Westlake Institute for Advanced Study, Hangzhou 310024, P. R. China.
    \\
    $^4$Key Laboratory of 3D Micro/nano Fabrication and Characterization of Zhejiang Province, School of Engineering, Westlake University, Hangzhou 310024, Zhejiang Province, P. R. China.
    \\
      $^5$ZJU-Hangzhou Global Scientific and Technological Innovation Center, School of Physics, Zhejiang University, Hangzhou 311215, China.
        \\
    $^6$Academy of Interdisciplinary Studies on Intelligent Molecules, Tianjin Key Laboratory of Structure and Performance for Functional Molecules, College of Chemistry, Tianjin Normal University, Tianjin, 300387, P. R. China.
        \\  
    $^7$Hangzhou Key Laboratory of Quantum Matters, School of Physics, Hangzhou Normal University, Hangzhou 311121, China.
        \\
    $^8$Material Science Center, Yangtze Delta Region Academy of Beijing Institute of Technology, Jiaxing 314011, China.
        \\
    $^9$Wuhan National High Magnetic Field Center, Huazhong University of Science and Technology, Wuhan, 430074, China.
    \\
    $^{10}$Centre for Quantum Physics, Key Laboratory of Advanced Optoelectronic Quantum Architecture and Measurement (MOE), School of Physics, Beijing Institute of Technology, Beijing 100081, China.
    \\
    $^{11}$Beijing Key Lab of Nanophotonics and Ultrafine Optoelectronic Systems, Beijing Institute of Technology, Beijing 100081, China.
    \\
    $^{12}$Center for Quantum Matter, Zhejiang University, Hangzhou 310058, China.
    \\
    These authors contributed equally$^\dag$: Zhefeng Lou, Yingjie Zhao, Zhihao Gong. \\$*$e-mail:	liwenbin@westlake.edu.cn (W.L.); zhengxiaorui@westlake.edu.cn (X.Z.); daodaohw@zju.edu.cn (H.W.); linxiao@westlake.edu.cn (X.L.)}}
    }
 
	\date{\today}
	\begin{abstract}
	\noindent  Two-dimensional (2D) materials with remarkable second-harmonic generation (SHG) hold promise for future on-chip nonlinear optics. Relevant materials with both giant SHG response and environmental stability are long-sought targets. Here, we demonstrate the enormous SHG from the phase engineering of a high-performance semiconductor, Bi$_2$O$_2$Se (BOS), under uniaxial strain. SHG signals captured in strained 20 nm-BOS films exceed those of NbOI$_2$ and NbOCl$_2$ of similar thickness by a factor of 10, and are four orders of magnitude higher than monolayer-MoS$_2$, resulting in a significant second-order nonlinear susceptibility on the order of 1 nm~V$^{-1}$. Intriguingly, the strain enables continuous adjustment of the ferroelectric phase transition across room temperature. Consequently, an exceptionally large tunability of SHG, approximately six orders of magnitude, is achieved through strain or thermal modulation. This colossal SHG, originating from the geometric phase of Bloch wave functions and coupled with sensitive tunability through multiple approaches in this air-stable 2D semiconductor, opens new possibilities for designing chip-scale, switchable nonlinear optical devices.
	\end{abstract}
\maketitle

\noindent\textbf{1. Introduction}

\noindent
Phase transitions can profoundly alter the physical or topological properties of two-dimensional (2D) materials~\cite{Jarillo-Herrero2021Science,ZhangX2017Nature,Appenzeller2019NM,LeeYH2017NP,Pasupathy2021Nature,YangH2015Science,YangXH2022AM,LiWB2021NRM}, the precise control of which through external stimuli holds significant technological potential across various fields, including electronics, optics, and catalysis\cite{LiWB2021NRM,ZhaiTY2018AFM,Chhowalla2015CSR,Lindenberg2019Science,MiaoF2024NM}. Among these, the ferroic phase transition is characterized by the loss of certain point-group symmetries, which holds promise for applications in high-speed non-volatile electronics~\cite{LiWB2021NRM,LiJ2021PNAS}.

Featuring a narrow bandgap (0.8 eV)\cite{ChenYulin2018Sci.adv.}, high electron mobility\cite{Peng2017Nat.Nanotechnol.,Peng2023Nat.Mater.}, and robust environmental stability\cite{li2DBi2O2SeEmerging2021}, the layered semiconductor Bi$_2$O$_2$Se (BOS) has become a highly competitive material for next-generation electronics~\cite{li2DBi2O2SeEmerging2021,tan2DFinFieldeffect2023,Peng2022Nat.Electron.,Peng2020Nat.Electron.} and broadband optoelectronics~\cite{khanControlledVaporSolid2019,tongSensitiveUltrabroadbandPhototransistor2019,fuUltrasensitive2DBi2O2Se2019,Peng2018Nat.Commun.,HuWeida2022Sci.Adv}. Our recent work reported local strain engineering of the paraelectric (PE) to ferroelectric (FE) phase transition in BOS~\cite{wu2023achieving}, highlighting BOS as a rare high-performance semiconductor that also exhibits an FE phase transition.


\begin{figure*}[!thb]
\includegraphics[width=18cm]{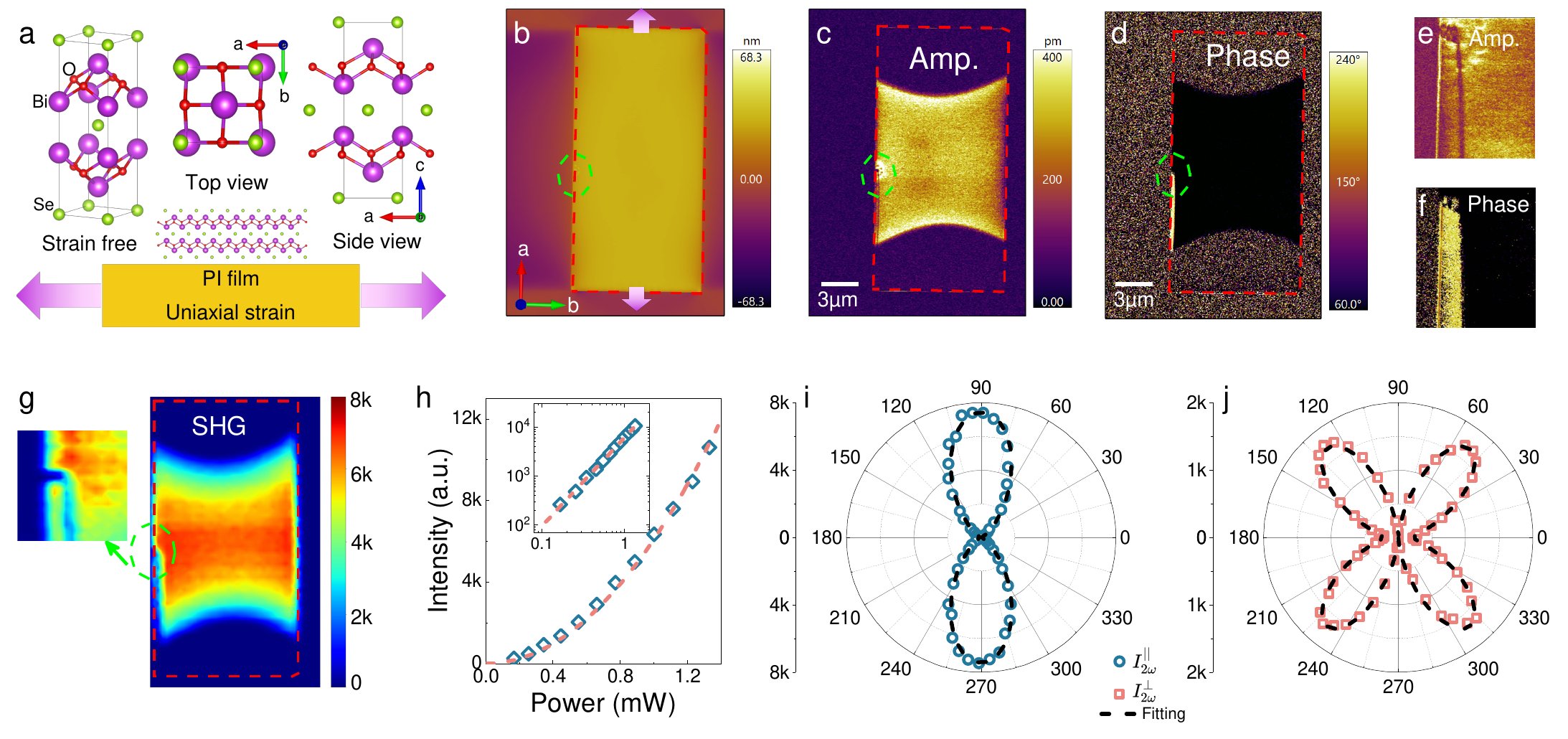}
		\renewcommand{\figurename}{\textbf{Figure}}
\caption{Ferroelectricity in strained BOS. a) Lattice structure of BOS subjected to in-plane uniaxial strain. 
b) Topography of a strained BOS film from atomic force microscopy. The arrows mark the strain applied along the $a$-axis (long edge), by sticking BOS to a flexible PI film as sketched in (a).
c) Mapping of the in-plane PFM amplitude on the film. d) Mapping of the corresponding phase. e,f) Enlarged images of PFM amplitude and phase, corresponding to the circled area in (b), (c), (d). It illustrates FE domains. g) Mapping of SHG intensity for the same device. The side figure is a magnified image showing FE domains.  h) SHG intensity as a function of excitation power. The dashed curve is a quadratic fit. The inset is the same data in a log-log plot. The dashed line is a guide to the eye. 
i,j) Angular dependence of polarized SHG signals. $I_\mathrm{2\omega}^{\parallel}$ and $I_\mathrm{2\omega}^{\perp}$ represent the measurements in the parallel and perpendicular modes, respectively. The dashed curves are fits to the  $mm2$ point group, as discussed in Note~S1, Supporting Information. The SHG measurements were performed with 1064~nm excitation. All the mapping was measured along the polar-axis as suggested by the maximum in (i). The angle of 90° corresponds to the strain direction. The red dashed boxes outline the specimen.}
\label{Fig1}
\end{figure*}

In this work, we discover that phase engineering enables giant second-harmonic generation (SHG) responses in BOS thin films through mechanical/thermal modulations around room temperature (RT). Strikingly, the maximum intensity observed in strained films exhibits a tenfold increase compared to that of NbOX$_2$ (X=Cl, I) of similar thickness at ambient conditions. The latter has recently received considerable attention due to its huge, thickness scalable SHG~\cite{GiantNbOI2,Andrew_T.S2023Nature,Huang_Bing2023NCNbOX2compression,pressNbOX2JACS}, as well as excellent spontaneous parametric down-conversion efficiency~\cite{Andrew_T.S2023Nature}. Importantly, the SHG signals in BOS exhibit exceptionally large strain tunability across FE transitions, evolving from below noise levels (i.e. no SHG in the PE phase) at ambient conditions to a value approximately $10^6$ times higher at the strain of $\varepsilon\approx4.1\%$. 
Similar behaviour can be achieved by thermal modulation, for which the FE Curie temperature ($T_\textrm{c}$) is continuously adjusted across RT by gradual strain tuning. This level of tunability dramatically exceeds that of other 2D systems subjected to various external stimuli~\cite{Liu2024NanoLett,ElectrialcontrolWSe2,ZhangX2017Nature,Zhang2021NatureElectronics}.  
Intriguingly, our first-principles calculations reveal that the geometric phases, such as shift vector, play a crucial role in the SHG response. The remarkable tunability of enormous SHG in BOS films via ferroic phase and geometric phase engineering opens up new possibilities for miniaturized solid-state optical applications, such as quantum light sources, optical modulation and switches, etc~\cite{review_2D,Zhang_2020_review_2D}.

\begin{figure*}[!thb]
	\includegraphics[width=16cm]{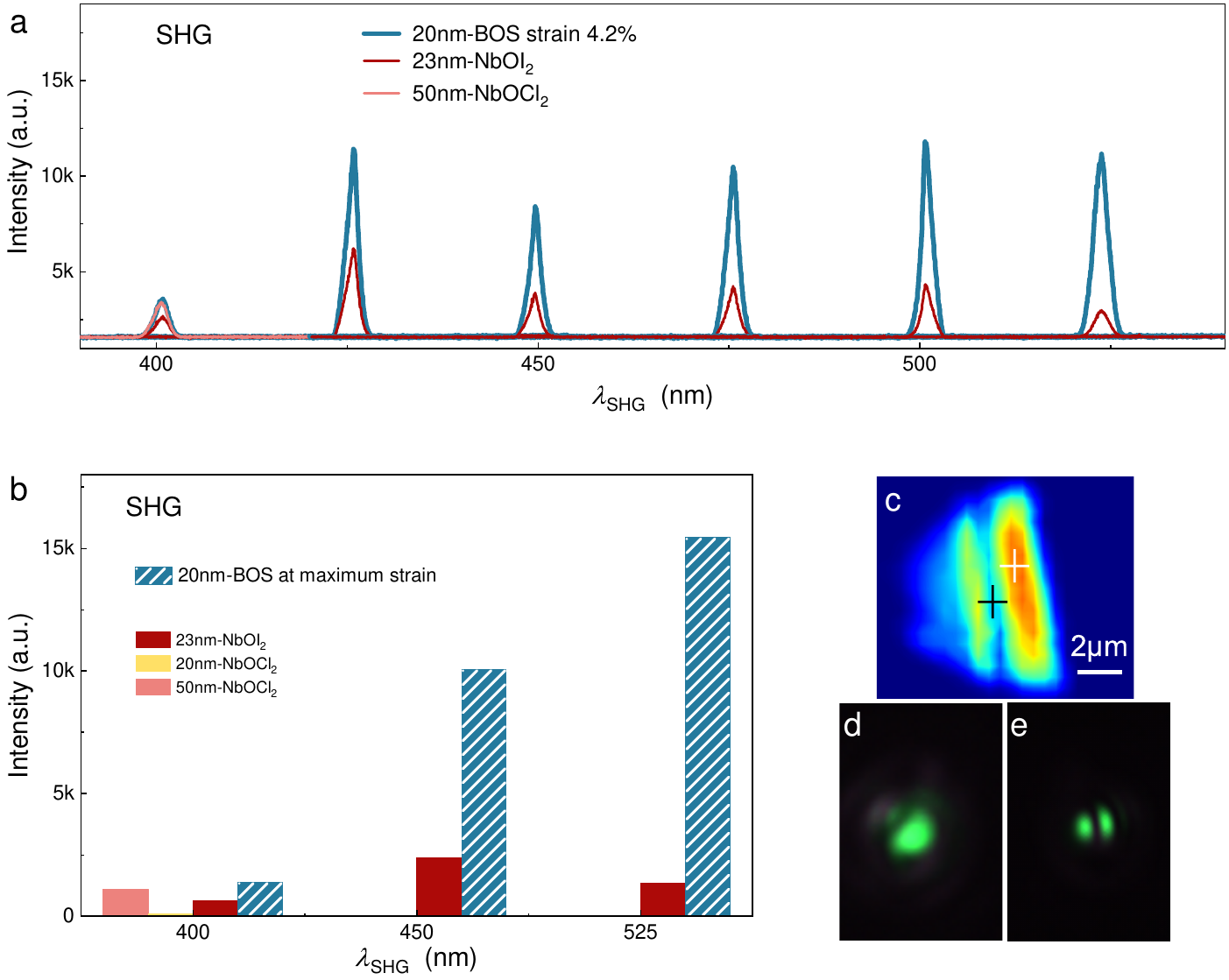}
	\renewcommand{\figurename}{\textbf{Figure}}
	\caption{Giant SHG in a strained BOS film with a thickness of 20~nm.  a) SHG intensity for strained BOS ($\varepsilon\approx 4.2\%$) at different wavelengths.  b) Maximum SHG of BOS at selective wavelengths ($\lambda_\textrm{SHG}=400, 450$ and $525$~nm) recorded prior to the film rupture under tension.  In (a) and (b), strain-free 23 nm-NbOI$_2$, 20 nm-NbOCl$_2$ and 50 nm-NbOCl$_2$ were employed as control samples. c) Mapping of SHG intensity near a domain boundary for strained BOS. d,e) Optical images of SHG emission located inside a domain (white `+' in (c))  and at the domain wall (black `+'), respectively. They are recorded in a confocal microscope with 1064~nm excitation. The spot size is about 1~$\mu$m$^2$.} 
	\label{Fig2}
\end{figure*}
~\\
\noindent\textbf{2. Results and Discussions}

\noindent\textbf{2.1 Strain induced ferroelectricity}

\noindent
BOS crystals possess inversion symmetry, adopting a tetragonal anti-ThCr$_2$Si$_2$ structure with the I4/mmm space group. Theoretical investigations suggest that the application of uniaxial tensile strain induces 
a relative in-plane shift between the [Se]$_\mathrm{n}^\mathrm{2n-}$ and [Bi$_2$O$_2$]$_\mathrm{n}^\mathrm{2n+}$ layers, as depicted in \textbf{Figure~\ref{Fig1}}a~\cite{WuMH2017,ZhuZiye2024J.Mater.Chem.C}. This disrupts inversion symmetry, leading to a FE transition~\cite{wu2023achieving}. In this study, the strain is achieved by attaching BOS films to stretchable polyimide (PI) substrates. A particular method is employed to enhance the adhesion between the BOS and PI interface, capable of sustaining strains on BOS films as large as $5\%$ see details in Experimental Section and Figure~S1, Supporting Information).

Figure~\ref{Fig1}b-f illustrates piezoelectric force microscopy (PFM) measurements on a strained BOS film. Figure~\ref{Fig1}b is the topography of the rectangular specimen with strain applied along the $a$-axis. Scanning images of in-plane PFM amplitude and phase are presented in Figure~\ref{Fig1}c-d, which reveal a foresail-like region with distinct contrast to the surroundings. This observation reflects the presence of a FE order in strained BOS~\cite{wu2023achieving}. The relatively weak out-of-plane PFM signal, as shown in Figure~S2, Supporting Information, further suggests an in-plane FE polarization, aligning well with density functional theory (DFT) calculations~\cite{WuMH2017,ZhuZiye2024J.Mater.Chem.C}.

Interestingly, a domain wall structure, extending from a point defect at the edge (enclosed by a circle), is detected in Figure~\ref{Fig1}c-d. The enlarged image in Figure~\ref{Fig1}e-f reveals two distinct regimes, separated by a line boundary, with identical PFM amplitude but 180° out of phase. Notably, this boundary does not correspond to a strain-induced fracture, as evidenced by the topography of the specimen.  Moreover, the domain walls exhibit adjustability based on the history of strain application, as seen in Figuer~S3, Supporting Information. All these indicate the presence of FE domain structure in strained BOS.

Figure~\ref{Fig1}g-j presents the primitive SHG measurement on the same device. In Figure~\ref{Fig1}g, the SHG intensity mapping revealed a foresail-like region of enhanced SHG responses as well as a domain wall structure, which appears to be identical to that detected by PFM, underscoring the intimacy between SHG and the FE phase. In Figure~\ref{Fig1}h, the SHG signals exhibit a typical scaling with the square of excitation power. 
Moreover, in Figure.~\ref{Fig1}i and j, the angle-dependence of polarization signals detected in the parallel-$I_\mathrm{2\omega}^{\parallel}$ and perpendicular-$I_\mathrm{2\omega}^{\perp}$ modes follows a twofold rotational symmetry from the predicted $mm2$ point group in the FE phase of strained BOS~\cite{wu2023achieving}. In Figure~\ref{Fig1}i, the maximum at 90° corresponds to the polar axis aligned with the strain ($a$-axis), as expected by DFT calculations~\cite{WuMH2017}. Details of the measurements are presented in Experimental Section; Figure~S4 and Note~S1, Supporting Information.

Note that the polar signals are observed only in the central foresail-like region, rather than across the whole sample. This observation aligns well with the simulated strain distribution on BOS films as seen in Figure~S5-S7, Supporting Informatin, where the strain transferred from the PI substrates shows moderate non-uniformity, with higher $\varepsilon$ in the middle and lower $\varepsilon$ on either side. The curved edges that separate the polar/non-polar regions in Figure~\ref{Fig1}c,d,g, correspond to $\varepsilon=\varepsilon_\mathrm{c}$~\cite{WuMH2017,wu2023achieving,LiWB2022JACS,ZhuZiye2024J.Mater.Chem.C}. 
It is important to emphasize that flexoelectricity induced by the non-uniform strain is negligible in contributing to the observed signals, as detailed in Figure~S6 and Note~S2, Supporting Information.

\begin{figure*}[!thb]
\includegraphics[width=18cm]{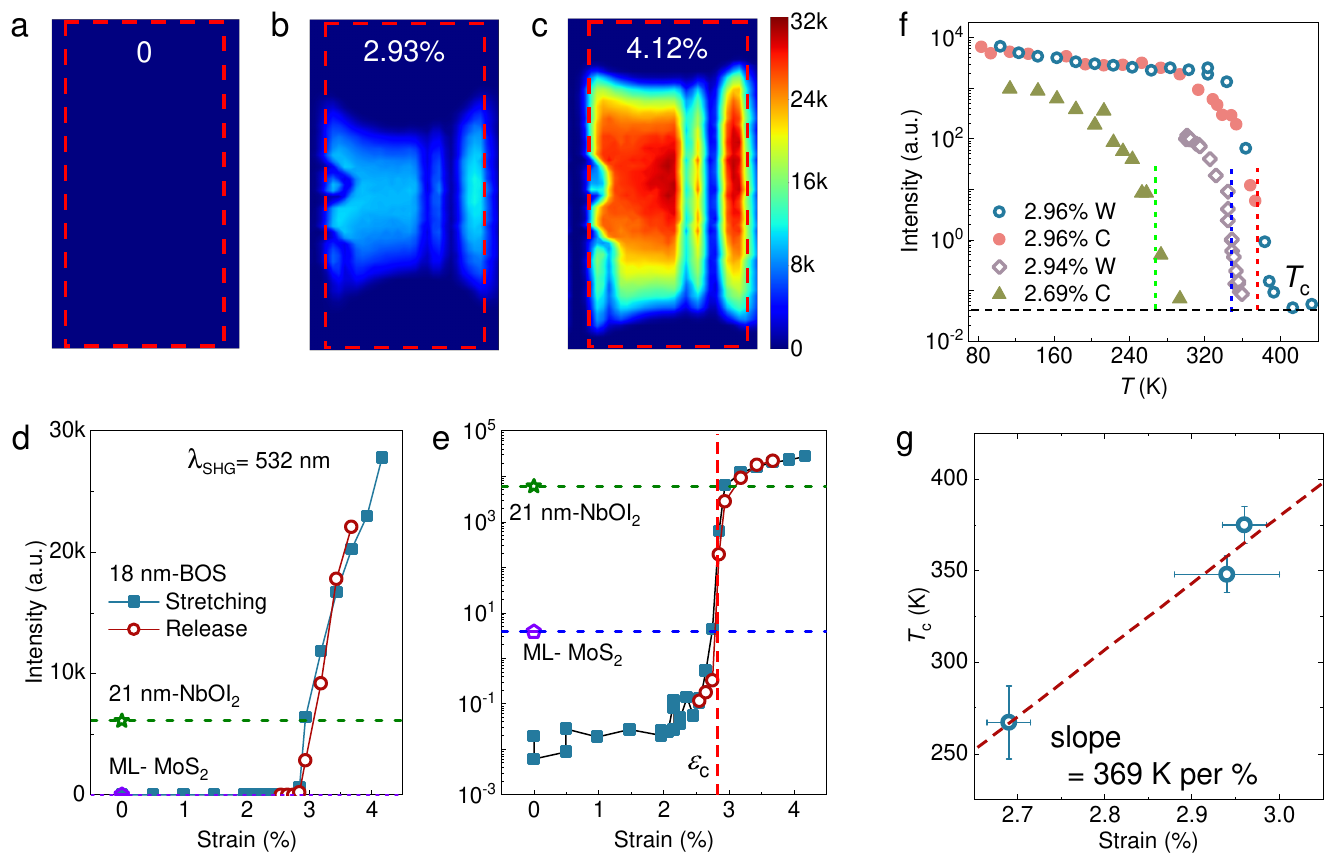}
\renewcommand{\figurename}{\textbf{Figure}}
  \caption{Multiple modulations of SHG.  a-c) SHG mapping for BOS at strains below (0), around (2.93\%) and above (4.12\%) $\varepsilon_\mathrm{c}$ of the FE transition. The dashed boxes outline the BOS film. The mapping was measured along the polar axis. Please refer to Figure~S11, Supporting Information, for the mapping along the non-polar axis, where SHG signals are negligible as strain varies. 
      d,e) Mechanical modulation of SHG intensity across the FE transition at RT. The data is plotted in linear and semi-log scale, respectively. 21~nm-NbOI$_2$ and ML-MoS$_2$ were employed as control samples.  f) Thermal modulation of SHG intensity at different strains around RT. The vertical dashed lines mark the FE Curie temperature, determined at the peak of the first derivative. `C'/`W' denotes cooling/warming. g) Strain modulation of $T_\mathrm{c}$. The dashed line is a linear fit. The error bars indicate uncertainties in determination of extracted values. The data was collected via 1064~nm excitation along the polar axis.
    }
\label{Fig3}
\end{figure*}

~\\
\noindent \textbf{2.2 Giant SHG} 

\noindent
\textbf{Figure~\ref{Fig2}}a illustrates the wavelength dependence of SHG intensity for a 20~nm-thick BOS film under substantial uniaxial strain ($\varepsilon\approx4.2\%$), with $\lambda_{\text{SHG}}$ spanning from 400~nm to 525~nm. It is compared with the data simultaneously obtained on a well-established 2D NbOX$_2$ system, specifically, 23~nm-NbOI$_2$ and 50~nm-NbOCl$_2$. Surprisingly, the strained BOS shows a higher SHG intensity than NbOX$_2$ films of similar thickness across the whole wavelength range measured. Consequently, we further stretch the BOS film and measure the SHG intensity at selected $\lambda_{\textrm{SHG}}$. In Figure~\ref{Fig2}b, the maximum SHG, recorded at $\varepsilon>5\%$ (just before the fracture of the sample), is compared with NbOX$_2$ at the characteristic wavelengths chosen in two seminal papers~\cite{GiantNbOI2,Andrew_T.S2023Nature}. The SHG response of 20~nm-BOS is approximately an order of magnitude larger than that of 23~nm-NbOI$_2$ at $\lambda_{\textrm{SHG}} = 525$ nm and than that of 20~nm-NbOCl$_2$ at $\lambda_{\textrm{SHG}} = 400$ nm (see Raman characterization in Figure~S8, Supporting Information). 


Given the second-order nonlinear susceptibility ($\chi^{(2)}$) of NbOI$_2$ and NbOCl$_2$ in Table~S1~\cite{GiantNbOI2,Andrew_T.S2023Nature}, Supporting Information, a significant $\chi^{(2)}$ on the order of 1~nm~V$^{-1}$ is inferred for strained BOS (see detailed calculations in Note~S3 and Figure~S9, S10, Supporting Information). It is more than one order of magnitude higher than that of LiNbO$_3$ (0.04~nm~V$^{-1}$)~\cite{Chekhova_M.V.2019PhysRevLett} and among the highest value of 2D materials, including transition metal dichalcogenides (TMDCs)~\cite{Zhao_Hui2013PRBMoS2500,LiuZhiwen2014ScientificReports}. While in TMDCs, the odd-even layer selectivity combined with the interlayer coupling impedes the thickness scalability of SHG response~\cite{review_2D,Zhang_2020_review_2D,Liu2017Adv.Mater.3R}.For example, the 3R-MoS$_2$ requires a much greater thickness ($\sim 700$~nm) to achieve a giant SHG response~\cite{Liu2024Science}. Similarly, comparable thicknesses are also needed for r-BN and BN nanotubes (BNNT) due to their small $\chi^{(2)}$ values($\chi^{(2)}_\mathrm{r-BN}\sim$0.03~nm~V$^{-1}$, $\chi^{(2)}_\mathrm{BNNT}\sim$0.05~nm~V$^{-1}$)~\cite{Liu_Kaihui2024AM,Liu2024Nat.Nanotechnol.}. 
Further comparison of $\chi^{(2)}$ with other systems is included in Table~S2, Supporting Information. In Figure~\ref{Fig2}d and e, we observe a remarkable SHG emission from the BOS films directly on the camera of a confocal microscope (see details in Experimental Section). The spots in two figures differ in shape, which is intrinsic to the SHG response to the location of domain bulk and domain walls, as seen in the SHG mapping of Figure~\ref{Fig2}c. Here, we achieve the first major outcome of present work, which may expand the functionality of the high-performance semiconductor BOS and motivate multiple applications in on-chip optical parametric oscillators and optical transistors~\cite{Safavi2022NC,Yilmaz2014J.LightwaveTechnol.,Signal_Processing_chip}.



\begin{figure*}[!thb]
\includegraphics[width=18cm]{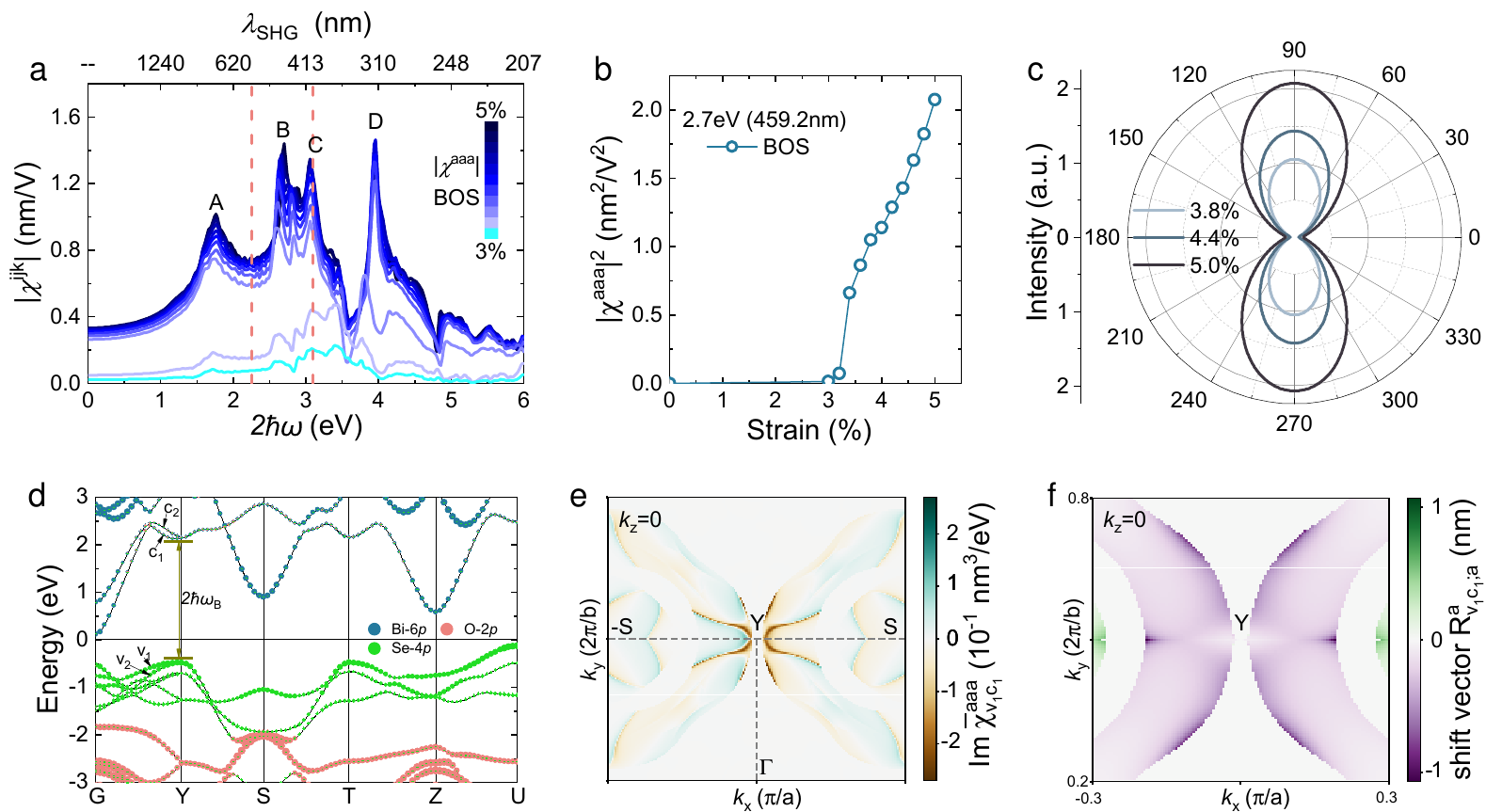}
\renewcommand{\figurename}{\textbf{Figure}}
\caption{First-principles calculations of SHG.  a) 
Calculated nonlinear susceptibilities $|\chi^{\text{aaa}}|$ of strained BOS, 
ranging from $\varepsilon=3\%$ (light blue) to 5\% (dark blue), at a strain step of 0.2\%. $a$ is the polar axis. Four peaks (A-D) of the SHG response are resolved.  Two vertical dashed lines enclose the frequency range of our measurement. 
 b) Calculated mechanical modulation of $|\chi^{\text{aaa}}|^2$ for BOS, with the SHG photon energy of 2.7~eV (around Peak-B). c) Angle-dependence of calculated total SHG intensities (see the definition in Note S1, Supporting Information) at different strains. 
 d) Band structure of BOS at $\varepsilon=5\%$.  The projections of Bi-$6p$, O-$2p$, and Se-$4p$ orbitals
 are highlighted with the size of circles indicating their weight in the bands.  
 e) ($k_\text{x}$, $k_\text{y}$) distribution of the representative ($v_1$, $c_1$)-band contributions to SHG,
 $\mathrm{Im}[\bar{\chi}_{\mathrm{v_1 c_1}}^{\text{aaa}}]$. 
 f) Corresponding distribution of the shift vector $R_{\mathrm{v_1 c_1;a}}^\text{a}$. The ($k_\text{x}$, $k_\text{y}$) heatmap is constructed at $k_\text{z}  = 0$ for Peak-B at $\varepsilon=5\%$.}
\label{Fig4}
\end{figure*}

~\\
\noindent\textbf{2.3 Multiple modulation of SHG response. } 

\noindent
In the following, we investigate the substantial mechanical and thermal modulation of SHG intensity on a 18~nm-BOS device. The measurements were performed at $\lambda_\textrm{SHG}=532~\text{nm}$ in a commercial instrument. \textbf{Figure~\ref{Fig3}}a-c depict the spatial distribution of SHG, demonstrating the variation of SHG intensity under three representative strains at RT. At $\varepsilon=0$, no SHG signal is detected, consistent with the PE phase at ambient conditions. This observation also rules out the possible SHG contribution from the PI-BOS or BOS-air interface that naturally lacks inversion symmetry.  While, at $\varepsilon\approx2.93\%$, discernible SHG appears in the central region of the sample, indicating the transition from the PE to FE phase. Upon further stretching the device to $\varepsilon\approx4.2\%$, the SHG intensity significantly strengthens. In Figure~\ref{Fig3}c, the SHG mapping also reveals multiple FE domain boundaries, primarily along the strain direction, which are absent in the optical image of Figure~S11a, Supporting Information. Since the polarizations of neighboring domains both align with the strain, these boundaries are generally neutral 180° domain walls. 

Figure~\ref{Fig3}d illustrates the evolution of SHG intensity as the strain varies, employing 21~nm-NbOI$_2$ and a CVD-grown 1H-MoS$_2$ monolayer (ML) as control samples. Notably, data from both stretching and release processes nearly collapse, indicating negligible relaxation during the strain transfer from PI to BOS even at $\varepsilon>4\%$. This underscores the robust adhesion at the interface, a crucial factor in maintaining structural integrity in strain engineering (see more in Figure~S7, Supporting Information).
This reversibility ensures a predictable SHG response within repetitive strain cycling, a critical feature for nonlinear optical modulators~\cite{ZhangX2017Nature}. In the figure, the critical strain  $\varepsilon_\mathrm{c} \approx 2.8(1)$ \% is clearly resolved, above which $I_\textrm{SHG}$ exhibits pronounced enhancement, surpassing that of NbOI$_2$ by a factor of 4.6 at $\varepsilon\approx4.2\%$, with no trend of saturation observed. Similar observations at other wavelengths are presented in Figure~S12, Supporting Information. The observed $\varepsilon_\mathrm{c}$ is close to the critical value of the PE-FE transition calculated in ref.~\citenum{ZhuZiye2024J.Mater.Chem.C}. It's noteworthy that the strain in BOS can be fixed using epoxy resin, and Figure~S13, Supporting Information demonstrates the environmental robustness of its SHG response over seven months. In contrast, NbOX$_2$ is less stable under ambient conditions~\cite{Andrew_T.S2023Nature,Fang2024ACSAppl.Mater.Interfaces}.

In Figure~\ref{Fig3}e, the same data, presented in a semi-log plot, highlights the exceptionally large tunability of $I_\textrm{SHG}$. $I_\textrm{SHG}$ evolves from the noise level (no SHG) at zero strain to a value six orders of magnitude higher at $\varepsilon\approx4.2\%$ (approximately $10^4$ times that of ML-MoS$_2$). In particular, the strain demonstrates a significant tunability of $I_\textrm{SHG}$ within a narrow strain range around $\varepsilon_\mathrm{c}$, i.e. a tunability of $I_\textrm{SHG}$ by a factor of $10^4$ within $\Delta\varepsilon\approx0.3\%$. 
Now, let's delve into the thermal modulation of SHG in BOS by adjusting the strain to bring $T_\textrm{c}$ of the FE transition close to RT. Figure~\ref{Fig3}f shows the $T$-evolution of SHG around RT at selective strains. As $T$ lowers, $I_\textrm{SHG}$ undergoes a steep rise corresponding to the PE to FE transition, resulting in a significant variation of five orders of magnitude. In Figure~\ref{Fig3}g, $T_\textrm{c}$ is roughly linear with strain, exhibiting sensitive tunability evolving from $T_\textrm{c}\approx267$~K at $\varepsilon\approx2.69\%$ to 375~K at $2.96\%$, corresponding to a large average slope $\mathrm{d}T_\textrm{c}/\mathrm{d}\varepsilon \approx 369$ K per \%. 
The dramatic tunability of both $I_\textrm{SHG}$ and $T_\textrm{c}$ in the vicinity of RT renders BOS highly valuable for advanced nonlinear optical signal processing based on mechano-optical or thermo-optical modulators~\cite{Xiao-MingChen2020NC,Chen_Ling2023Angewandte,Wang2016NaturePhotonics}. These results represent the second major outcome of this work.

~\\
\noindent\textbf{2.4 Microscopic origin and geometric aspects}

\noindent
Our experimental findings strongly indicate a close relationship between the pronounced SHG response and the FE phase transition in strain-modulated BOS. To gain a microscopic understanding of the SHG response, we performed first-principles calculations of the SHG susceptibility $\chi^\text{ijk}$. \textbf{Figure~\ref{Fig4}}a shows the calculated spectrum of the major component ($|\chi^\text{aaa}|$) for BOS under various strains ($\varepsilon>\varepsilon_\mathrm{c}$). Its comparison with the minor components is shown in Figure~S14 and S15, Supporting Information. The spectrum exhibits significant values across a broad wavelength range, with four distinct peaks (labeled A-D) between 310~nm and 830~nm.
We propose that our experimental SHG response originates from the spectral region around Peak-B, highlighted by the vertical lines in Figure~\ref{Fig4}a. In this region, $|\chi^\text{aaa}|$ approaches a value of approximately 1 nm~V$^{-1}$, in close agreement with our experimental results.
The strain dependence of $|\chi^\mathrm{aaa}|^2$ ($\sim I_\mathrm{2\omega}$, see Note~S3, Supporting Information) at 459.2~nm (around Peak-B) and the planar angular anisotropy of total SHG intensity are presented in Figure~\ref{Fig4}b and c, respectively. Above $\varepsilon_\textrm{c}$ ($\sim 3\%$), a significant enhancement in $|\chi^{aaa}|^2$ and a two-fold rotational symmetry, preserved for various strains, are observed. These theoretical predictions agree quantitatively with our experimental findings in Figure~S12, Supporting Information.


Figure~\ref{Fig4}d illustrates the calculated band structure of BOS at $\varepsilon=5\%$. The valence bands near the gap ($v_1$ and $v_2$) arise from Se-4$p$ orbitals, while the conduction bands ($c_1$ and $c_2$) come from Bi-6$p$ orbitals. Deeper valence bands are attributed to O-2$p$ orbitals. The double-photon resonant transition between $v_1$ and $c_1$ around the high-symmetry $Y$-point coincides with Peak-B in the spectrum. The relatively less dispersive bands around $Y$ leads to higher density of states (DOS) and increased absorbance near the gap in Figure~S16, Supporting Information. This consequently triggers strong nonlinear response around $Y$-point with the major contribution from the ($v_1$, $c_1$) transition, as seen in Figure~\ref{Fig4}e, which illustrates the distribution of SHG susceptibility within the ($k_\mathrm{x}$, $k_\mathrm{y}$)-plane at $k_\mathrm{z} = 0$ at $\varepsilon=5\%$ (see more discussions associated with other strains in Figure~S17-S20, Supporting Information).

As shown in Figure~S21, Supporting Information, the significant strain tunability of SHG in BOS primarily arises from the strain sensitivity of the ``shift" component ($\chi^\mathrm{aaa}_\text{shift}$) of the total SHG susceptibility. This term dominates the contribution over the ``triple" and ``inject" components and is associated with the geometric quantity termed the shift vector $R_\mathrm{nm;a}^\text{a}$, which elucidates the spatial displacement of charge centers along the $a$-direction upon $a$-polarized photon absorption with $n,m$ the band index.
Figure~\ref{Fig4}f presents the distribution of $R_\mathrm{v_1 c_1;a}^\mathrm{a}$ around $Y$-point, corresponding to the region of the most pronounced SHG susceptibility in Figure~\ref{Fig4}e. The similar profiles in two figures highly suggest the dominant role of geometric shift vector in inducing the SHG susceptibility around Peak-B.

Notably, $R_\mathrm{nm;a}^\text{a}$ is related to the electronic part of FE polarization since both are proportional to the intra-band Berry connection $A_\mathrm{n}^\mathrm{a}$ (see details in Note~S4, Supporting Information). As depicted in Figure~S22, Supporting Information, a finite distribution of $R_\mathrm{v_1 c_1;a}^\mathrm{a}$ around $Y$-point emerges once the strain-induced FE distortion disrupt the inversion symmetry at $\varepsilon>3\%$. Its overall magnitude increases with strain, correlated with the substantial enhancement in the ($v_1$, $c_1$)-band contribution to the SHG susceptibility as detailed in Figure~S18 and S19, Supporting Information. Therefore, understanding the evolution of the electronic geometric phase are crucial for explaining the giant SHG observed in BOS during the FE phase transition.

~\\
\noindent\textbf{3. Conclusion}

\noindent
Our work demonstrates giant and exceptionally tunable SHG in BOS nanosheets through the strain or thermal modulation of FE phase transition. A deep connection between the giant SHG, the FE phase transition, and the geometric phase of the electrons is also revealed, imparting a new meaning to the concept of phase engineering by encompassing both the FE phase and the geometric phase. The unprecedented tunability of the SHG responses in BOS near RT, coupled with the environmental stability and other attractive electronic and optoelectronic properties of BOS~\cite{li2DBi2O2SeEmerging2021,tan2DFinFieldeffect2023,khanControlledVaporSolid2019,tongSensitiveUltrabroadbandPhototransistor2019,fuUltrasensitive2DBi2O2Se2019}, may open the route towards a variety of applications involving highly sensitive, switchable nonlinear optics~\cite{Huang_Bing2023NCNbOX2compression,Xiao-MingChen2020NC,Chen_Ling2023Angewandte}. The interplay between the geometric phases of electrons and ferroic phase transitions may also open new avenues to explore emerging physics as well as unexpected functionalities in conventional and novel materials, which is of paramount interest for future research.

~\\


\noindent\textbf{4. Experimental Section}

\noindent
\textit{Sample Growth:} Inclined BOS nanosheets were synthesized via the physical vapor deposition (PVD) method, utilizing Bi$_2$O$_3$ powder (99.99\%, Aladdin) and BOS single crystals as precursors. These precursors are placed at the heating center of a horizontal oven and heated to 740 $^{\circ}\text{C}$. High purity argon was employed as the gas agent to deliver the precursors onto fluorophlogopite mica at a flow rate of $100-200$ sccm and pressure of 380 Torr. The mica was positioned at $10-13$ cm downstream of the precursors. 
This reaction lasted for 10 minutes, after which the oven was naturally cooled down to room temperature, resulting in the formation of inclined BOS nanosheets.

Bulk NbOX$_2$ single crystals were synthesized using the chemical vapor transport (CVT) method as previously reported~\cite{GiantNbOI2,Andrew_T.S2023Nature}. For NbOI$_2$, high-purity (5N) Nb and Nb$_2$O$_5$ powders were mixed together with I$_2$ flakes in a molar ratio of $3:1:5$ as source materials. For NbOCl$_2$, high purity Nb, Nb$_2$O$_5$ and NbCl$_5$ (5N) powders were mixed in stoichiometric ratio. They were then loaded into a 22 cm long quartz ampoule and sealed under high vacuum ($<10^{-2}$ Pa). The ampoule was placed in a dual-zone horizontal oven. The oven was then heated from RT to 873 K over 1 day and held at this temperature for 5 days. Subsequently, one end of the ampoule containing raw materials was cooled to 583 K at a rate of 1.2 K/h, while the opposite end was cooled to 513 K at a rate of 1.5 K/h. The resulting single crystals were millimetres in size with a lustrous appearance. 
NbOX$_2$ single crystals were stored in an N$_2$-filled glove box to prevent deliquescence. Thin films of NbOCl$_2$ and NbOI$_2$ were obtained by mechanical exfoliation of single crystals onto SiO$_2$/Si and quartz substrates, respectively.

1H-ML MoS$_2$ were grown on sapphire (0001) substrates by  low pressure chemical vapor deposition (CVD) under S-rich conditions~\cite{XingranWangNN2021}. Elemental S powder (99.95\%, Aladdin) and MoO$_3$ powder (99.95\%, Alfa Aesar) were used as the sulfur and molybdenum sources, respectively. Prior to growth, the substrates were subjected to low pressure annealing at 1000 $^\circ$C for 4 hours under a 400 sccm Ar and 100 sccm O$_2$ gas flow. 

~\\

\noindent\textit{Strain engineering:} The devices for strain engineering were fabricated by transferring BOS nanosheets onto a stretchable polyimide (PI) film. To ensure robust adhesion at the interface between BOS and PI, a specific process was employed as follows: First of all, 15\% polyamic acid solution was spin-coated onto the surface of PI substrate (50 $\mu$m) at a speed of 3000 rpm for 1 min, followed by baking at 70 $^\circ$C for 24 h to remove the dimethylacetamide solvent. Second, inclined BOS nanosheets were transferred onto the substrate prepared in the first step using a tip-assisted transfer method~\cite{hongInclined}. Finally, the device was heated at 180 $^\circ$C for 30 min to polymerize the polyamic acid into PI. During the polymerization process, a strong adhesion was formed at the interface, ensuring effective strain transfer from PI substrates to BOS films.  The in-plane uniaxial strain was applied using a homemade stretching apparatus, as shown in Figure~S1, Supporting Information.

This method enables much larger strains on the nanosheets compared to techniques that directly transfer 2D nanosheets onto flexible substrates~\cite{Liu2024NanoLett,StrainMLMoS2,strainMoS2MoSe2,StrainSTO}. As in our experiment, we did not observe any strain relaxation phenomena, such as wrinkling~\cite{Tang_Zikang2019wrinkling} or slippage~\cite{Zhang_Qian2018ACS}, even when the strain on BOS exceeded 5\%. It appears that the maximum strain is only limited by the quality of target films as evidenced by the film rupture under strain, rather than the interaction strength between BOS and substrates. 
This technique may have broader applications in the field of strain engineering for other 2D systems. We note that a PVA-assisted strain transfer technique could also achieve considerable strain on MoS$_2$ by bending, which differs from our approach. For further details, please refer to ref.~\citenum{YuanLiu2020NCefficient}.

~\\
\noindent\textit{PFM measurements:} PFM measurements were conducted using a vector-PFM modulus in a Jupiter XR atomic force microscope. Pr/Ir-coated tips with a radius of $<25$ nm and a spring constant of about 2 N/m were utilized as probes. The out-of-plane PFM signal was recorded by detecting the deflection of cantilever, while the in-plane signal was captured by the torsion of cantilever, which correlates with the in-plane deformation. Throughout the measurements, the cantilever was positioned perpendicular to the polar axis.

~\\
\noindent\textit{SHG measurements:} The measurements in Figure~\ref{Fig1}, \ref{Fig2}c,d and \ref{Fig3} were conducted using the reflection geometry of a confocal microscope (WITec, Alpha300RAS) equipped with a 1064 nm laser excitation source (NPI Rainbow 1064 OEM) with a pulse duration of 8.4~ps and a repetition rate of 50~MHz. The excitation beam was focused onto the sample with a spot size of approximately $1~\mu$m$^2$ using a $\times100$ objective lens with a numerical aperture (NA) of 0.75 for measurements at RT. To improve the resolution for small SHG signals in BOS under low strains and ML-MoS$_2$, both laser power and integration time were increased. For comparative analysis, the SHG signals in Figure~\ref{Fig3}d and e were normalized to a value calculated assuming $P=1.38$~mW and $\tau=0.1$~s, since SHG intensity exhibits a linear relationship with the square of power and integration time ($\bar I_\textrm{SHG}=I_\textrm{SHG}(P,\tau)\frac{1.38~\text{mW}\cdot0.1~\text{s}}{P^2\cdot \tau}$). 
This normalization ensures the accurate comparison across different experimental conditions. In Figure~\ref{Fig3}f, $T$-dependent SHG signals were collected using a continuous-flow optical microscopy cryostat (Linkam THM600 stage) with a $\times50$ objective lens (NA = 0.55). Similar to previous measurements, the intensity was normalized to facilitate comparative analysis across different temperatures. $T_{\mathrm{c}}$ and $\varepsilon_{\mathrm{c}}$ were determined from the peak of the first derivative of the semi-log plots in Figure~\ref{Fig3}e,f. In Figure~\ref{Fig2}c,d, the optical image  are recorded by a CMOS camera (Imaging Source) with an excitation power ($P$) of 9.7~mW.

%
The measurements in Figure~\ref{Fig2}a and b were carried out in a homemade system with a reflection geometry. Here, a femtosecond laser (Light Conversion, PHAROS PH2-10W) was utilized to pump the optical parametric amplifier (Light Conversion, ORPHEUS-F, pulse duration of 180~fs, repetition rate of 10~kHz) to generate laser beams of wavelengths ranging from 800~nm to 1100~nm in steps of 50 nm. The incident beam was directed through an $\times$40 objective lens (Olympus, RMS40X-PF,  NA=0.75) and focused onto the samples with a spot radius of approximately 1.5~$\mu$m. Due to the reflection geometry, the outgoing SHG signal was captured by the same objective and analyzed by an imaging spectrometer (Princeton Instruments, IsoPlane SCT 320). During the measurements, the excitation laser power was carefully maintained at 5 $\mu$W, as monitored by a silicon photodiode power sensor (Thorlabs, S130C). 


~\\
\noindent\textit{First-principles calculations:} The Kohn-Sham electronic structure of optimized crystal configuration was achieved by performing density functional theory (DFT) calculations~\cite{hohenberg1964PR, kohn_1965PR} using  
the Vienna \textit{Ab initio} Simulation Package (VASP)~\cite{kresse1993PRB, kresse1996CMS} code. The exchange-correlation energy of valence electrons was treated by the generalized gradient approximated (GGA) functionals~\cite{becke1988PRA,langreth1983PRB} of Perdew-Burke-Ernzerhof (PBE) type parameterization~\cite{perdew1996PRL}. An energy cutoff for the plane-wave basis was set to $400 \text{ eV}$, with a $9\times 9\times 3$ Monkhorst-Pack \textit{k}-point sampling. Convergence was reached when residual forces in ion relaxation were below $10^{-3}~\text{eV}/$\AA ~and the energy difference between sequential steps of electronic self-consistent field calculation was less than $10^{-6}\text{ eV}$. 
For the efficient evaluation of SHG susceptibility, the Hamiltonian was obtained using the Wannier tight-binding method~\cite{marzari2012RMP,ibanez-azpiroz2018PRB,garcia2023PRB} implemented in the Wannier90 code~\cite{pizzi2020JPCM}.
In the Wannier tight-binding calculations, a dense \textit{k}-point 
mesh of $100\times 100\times 40$ for converged integrals of the Brillouin zone and an energy width of $\eta = 0.025\text{ eV}$ for smearing Dirac $\delta$-functions
were adopted.

~\\
\noindent \textbf{Supporting Information}

\noindent Supporting Information is available from the author.

~\\



\noindent\textbf{Acknowledgments}

\noindent This research is supported by ``Pioneer'' and ``Leading Goose'' R$\&$D Program of Zhejiang under Grant 2024SDXHDX0007, Zhejiang Provincial Natural Science Foundation of China for Distinguished Young Scholars under Grant No. LR23A040001 and the Research Center for Industries of the Future (RCIF) at Westlake University under Award No. WU2023C009 and WU2022C024. W.L. acknowledges the support by NSFC under Award No. 62374136. H.W. acknowledges the support provided by NSCF under Grant No. 12304049. X.Z. acknowledges the support by Westlake Institute For OptoElectronics. We thank the support provided by Lin Liu from Instrumentation and Service Center for Physical Sciences (ISCPS) and Zhong Chen for Molecular Sciences (ISCMS) at Westlake University.

~\\
\noindent\textbf{Author contributions}

\noindent  Z.L., Y.Z. and Z.G. contributed equally to this work. Z.L. fabricated the device and performed PFM measurements with the assistance of M.W., H.Q., H.Z., Z.X. and L.L.. Z.L. and Y.Z. did optical measurements supervised by X.Z. and X.L.. Z.G. and Z.Z. did DFT calculations supervised by W.L. and H.W..  T.W. grew BOS nanosheets. J.S., W.K. and S.X. offered high-quality ML-MoS$_2$. J.W. offered NbOI$_2$ single crystals. Z.W. offered NbOCl$_2$ single crystals. Z.L., Z.G. and X.L. prepared the figures. W.L., H.W. and X.L. wrote the paper with the inputs from Z.L. and Z.G.. X.L. led the project. All authors contributed to the discussion.

~\\

\noindent\textbf{Conflict of Interest} 

\noindent The authors declare no conflict of interest. 

~\\

\noindent\textbf{Data Availability Statement} 

\noindent The data that support the findings of this study are included in this article and its supplementary information file and are available from the corresponding author upon reasonable request.
~\\

\noindent\textbf{Ketywords}

\noindent bismuth oxyselenide, second harmonic generation, ferroelectric transition, strain engineering

\normalem
\bibliography{BOSSHG}
\bibliographystyle{naturemag}



\end{document}